\documentclass[11pt,a4paper]{article}

\usepackage[utf8]{inputenc}
\usepackage[T1]{fontenc}
\usepackage[margin=1.1in]{geometry} 
\usepackage{charter} 
\usepackage{cite}
\usepackage{amsmath,amssymb,amsfonts}
\usepackage{graphicx}
\usepackage{tabularx}
\usepackage{booktabs} 
\usepackage{url}
\usepackage{xcolor}
\usepackage{authblk}
\usepackage{sectsty} 

\definecolor{arxivblue}{RGB}{42, 82, 190}
\sectionfont{\color{arxivblue}\sectionrule{0pt}{0pt}{-5pt}{0.8pt}} 
\subsectionfont{\color{arxivblue}}

\title{\vspace{-1cm}\Large \textbf{How Open is Open TTS?} \\ 
\large \textit{A Practical Evaluation of Open Source TTS Tools}}

\author[1]{\textbf{Teodora Răgman}}
\author[2]{\textbf{Adrian Bogdan Stânea}}
\author[1]{\textbf{Horia Cucu}}
\author[1,2]{\textbf{Adriana Stan}}

\affil[1]{POLITEHNICA Bucharest, Romania}
\affil[2]{Technical University of Cluj-Napoca, Romania}
\affil[ ]{\small \texttt{adriana.stan@com.utcluj.ro}}

\date{Published in IEEE Access \\ \url{https://ieeexplore.ieee.org/document/11269795}} 

\begin{document}

\maketitle
\begin{abstract}
   
    Open-source text-to-speech (TTS) frameworks have emerged as highly adaptable platforms for developing speech synthesis systems across a wide range of languages. However, their applicability is not uniform -- particularly when the target language is under-resourced or when computational resources are constrained. In this study, we systematically assess the feasibility of building novel TTS models using four widely adopted open-source architectures: FastPitch, VITS, Grad-TTS, and Matcha-TTS. Our evaluation spans multiple dimensions, including qualitative aspects such as ease of installation, dataset preparation, and hardware requirements, as well as quantitative assessments of synthesis quality for Romanian. We employ both objective metrics and subjective listening tests to evaluate intelligibility, speaker similarity, and naturalness of the generated speech. The results reveal significant challenges in tool chain setup, data preprocessing, and computational efficiency, which can hinder adoption in low-resource contexts. By grounding the analysis in reproducible protocols and accessible evaluation criteria, this work aims to inform best practices and promote more inclusive, language-diverse TTS development.
    All information needed to reproduce this study (i.e. code and data) are available in our git repository~\footnote{\url{https://gitlab.com/opentts_ragman/OpenTTS}}.

\end{abstract}

\maketitle

\section{Introduction}

Large-capacity deep neural networks play a crucial role in advancing speech generation because they can model the complex, high-dimensional relationships between linguistic, acoustic, and prosodic features required for natural-sounding speech. Speech is not only about producing correct words—it involves capturing subtle variations in tone, rhythm, emotion, and speaker identity. Larger models, with their greater number of parameters and deeper architectures, have the representational power to learn these intricate patterns from vast datasets, enabling them to produce more human-like intonation, smoother transitions between phonemes, and more consistent voice quality. This capacity also allows for better generalization across diverse speakers, languages, and styles, making them suitable for applications like conversational AI, dubbing, assistive technologies, and creative content generation. In essence, the size and depth of these networks directly translate into their ability to synthesize speech that is both intelligible and expressive.

Despite their impressive capabilities, large-capacity deep neural networks present notable challenges when applied to \textit{underserved languages}.\footnote{We note here that underserved does not equate to low-resource. Underserved means that speech and text data are available, but the research and commercial communities have not yet focused on developing the technology for the respective language.} These networks typically require massive amounts of high-quality, annotated speech data to achieve their full potential. As a result, the models risk overfitting to limited datasets or inheriting biases from dominant languages present in multilingual training corpora, leading to mispronunciations, unnatural prosody, or the erosion of language-specific nuances. Furthermore, training and fine-tuning such large networks demand substantial computational resources, which can be cost-prohibitive for research teams or communities working on less spoken languages. This creates a techno-ical gap where high-performing speech generation is disproportionately accessible for well-resourced languages, while those with smaller speaker populations remain underserved in generative AI-driven applications.

\textbf{Problem}. While open-source text-to-speech (TTS) frameworks offer substantial flexibility for building speech synthesis systems, their performance and usability are not consistent across all languages or computational environments. Underserved languages, such as Romanian, face a dual challenge: limited availability of high-quality training data and the high computational demands of state-of-the-art architectures.

\textbf{Motivation}. As generative speech technologies mature, ensuring equitable access to high-quality synthesis capabilities across diverse language families becomes increasingly important. Open-source models hold the promise of lowering entry barriers for research and development, but practical obstacles in installation, dataset preparation, and compute efficiency can prevent their effective adoption for less widely spoken languages. A systematic, comparative assessment is needed to identify these obstacles and to guide practitioners toward workable, reproducible solutions.

Unlike previous studies that primarily compare TTS models based on reported results, our study takes a hands-on, experimental approach to evaluating open-source systems. While
current research focuses on speech quality, controllability or architecture improvements, many surveys lack practical validation.
For instance, \cite{tan2021surveyneuralspeechsynthesis} provides an overview of speech synthesis techniques and open-source implementations, but relies solely on documented results, making direct comparisons difficult due to different training and evaluation setups. Other studies focus on specific aspects, such as controllable speech synthesis~\cite{xie2024controllablespeechsynthesisera} or diffusion-based models~\cite{zhang2023surveyaudiodiffusionmodels}. Although valuable, these works evaluate isolated categories of models, with performance results mainly derived from English-based checkpoints and little exploration of underserved languages. A study with a related approach to this one is that of Patole et al.~\cite{Patole2021} which evaluates the selected TTS models using MOS (Mean Opinion Score). However, details on the experimental setup, including the evaluation language, are not fully provided.

\textbf{Contributions}. This study evaluates the feasibility of developing novel TTS systems for Romanian using four popular open-source architectures. We conduct qualitative assessments in terms of ease of installation, dataset preparation complexity, as well as quantitative evaluations of intelligibility, speaker similarity, and naturalness using objective and subjective metrics. Our findings highlight key technical and procedural bottlenecks while outlining best practices for low-resource TTS development. By grounding the analysis in reproducible workflows and accessible evaluation criteria, we aim to support more inclusive, scalable, and language-diverse TTS research.

\section{The speech generation landscape}

Nowadays, generative speech or more traditionally speech synthesis is a stack of complementary advances: neural codecs that compress and represent waveforms, acoustic and waveform generators based on diffusion/flow/normalizing-flow/variational families, and highly engineered commercial systems that package these models into APIs and products. Together they enable near-human naturalness, low-latency serving, and flexible voice-style control, but they also bring compute, data and safety trade-offs.

Neural audio codecs such as SoundStream~\cite{zeghidour2021soundstream}, EnCodec~\cite{defossez2023encodec} and follow-ups learn compact, high-fidelity latent representations of raw waveforms using end-to-end encoder-decoder architectures with vector quantization and adversarial or perceptual losses. These codecs serve two important roles: (i) they let generative models work in a discrete or low-rate latent space — drastically reducing compute and sample time compared with operating at raw waveform resolution — and (ii) they enable codec-level editing, compression, and streaming for production use. SoundStream demonstrated real-time, low-bitrate speech and music compression with human-rated quality, and Meta’s EnCodec is now widely used as the latent front-end for downstream audio generation. For example, ParlerTTS~\cite{lyth2024parlertts} combines descriptive prompts with the Descript Audio Codec~\cite{kumar2023highfidelityaudiocompressionimproved} to elicit controllable, expressive text-to-speech synthesis. 

In terms of more standard approaches to the task of speech synthesis or generation, 
recent work has shown that diffusion~\cite{ICML-2015-Sohl-DicksteinW} and normalising flow~\cite{pmlr-v37-rezende15} methods produce highly natural mel-spectrograms or latents when used as decoders. Some of the most notable architectures are: 
WaveGrad~\cite{chen2021wavegrad}, Guided-TTS~\cite{kim2021guidedtts}, FastDiff~\cite{huang2022fastdiff}, DiffWave~\cite{kong2021diffwave} or Grad-TTS~\cite{popov2021gradttsdiffusionprobabilisticmodel} which introduced score-based diffusion decoders for TTS, showing strong subjective quality while offering explicit control over the quality/speed tradeoff; Flowtron~\cite{valle2020flowtron}, FlowTTS~\cite{miao2020flowtts}, PortaSpeech~\cite{ren2021portaspeech}, Glow-TTS~\cite{kim2020glow}, and 
Matcha-TTS~\cite{mehta2024matcha} applied conditional flow matching, an ODE-based (Ordinary Differential Equation) approach related to rectified flows, to achieve fast sampling with competitive quality; and flow/ODE or score-based decoders have been combined with alignment modules to remove dependence on external aligners. These methods have become central in recent papers and codebases because they balance probabilistic expressivity (to model variability and prosody) with mechanisms to speed up inference (fewer iterative steps, distillation, ODE samplers). 

A previous generation of speech synthesis architecture, yet one that still achieves high quality output is that based also on variational inference.
For example VITS~\cite{kim2021vits} combines variational inference, normalizing flows and adversarial training to produce single-stage, end-to-end TTS systems that avoid separate spectrogram-to-waveform stages while achieving state-of-the-art naturalness. Such models integrate alignment, acoustic modelling and waveform synthesis into one network, simplifying pipelines and improving prosody and expressiveness when trained on adequate data. 

We note here that for most of these systems, the authors have also released their source code, and associated inference and/or training recipes. This, at least hypothetically, should make the architectures readily available for direct use, reproduction or adaptation to other domains, speakers or languages. 
As a result, major companies operationalize all these research advances into production-grade APIs for voice cloning, multi-style TTS, and speech-to-speech. The research allows them to obtain ultra-realistic, low-latency voice synthesis and enterprise SDKs for dozens of languages. 


\textbf{Ethics, safety and accessibility concerns}.
The same systems that enable high-fidelity cloning and style transfer raise misuse risks (unauthorized voice cloning, deepfake audio). Commercial vendors increasingly add watermarking/detectors and human-in-the-loop policies, but robust provenance and detection remain active research and policy areas. Moreover, the compute/data demands of SotA models can widen access gaps for underserved languages and smaller labs — reinforcing the need for efficient codecs, transfer learning, limited-resource adaptation, and community datasets.


\section{Methodology}

Evaluating the complete landscape of open source TTS architectures is definitely unfeasible. We, therefore, select four of the most representative and referenced tools in the recent scientific literature. The tools have shown their flexibility across application domains and languages, and constitute a relevant starting point for the type of analysis we aim to perform. For each system, we used the official or widely-adopted implementations in an effort to ensure reproducibility and consistency across experiments, as well as for future reference. 

This section details the selected models' architecture, the speech data used for training and evaluation, as well as the text processing module and training environment setup. 

\subsection{TTS systems selection}

\textbf{FastPitch}\footnote{\url{https://github.com/NVIDIA/DeepLearningExamples/tree/master/PyTorch/SpeechSynthesis/FastPitch}}~\cite{fastpitch} is a fully non-autoregressive text-to-speech model designed for fast, high-quality speech synthesis with explicit prosody control. It builds on the FastSpeech 2 framework by incorporating a feed-forward Transformer-based encoder–decoder that predicts Mel-spectrograms in parallel, eliminating the sequential dependencies of autoregressive models and thus enabling low-latency synthesis. A central innovation is the explicit prediction of the fundamental frequency (F0) contour at the phoneme level, which is then added to the encoder output as a conditioning signal for the decoder. This mechanism allows fine-grained control over pitch, enabling prosody manipulation without retraining. The architecture uses a duration predictor to align phoneme sequences with Mel frames, removing the need for attention-based aligners and improving robustness against alignment errors. In evaluations, FastPitch demonstrated substantial speedups compared to autoregressive baselines while maintaining competitive naturalness, making it well suited for large-scale and interactive applications.


\textbf{VITS}\footnote{\url{https://github.com/jaywalnut310/vits}}~\cite{kim2021vits} is an end-to-end text-to-speech model that unifies text–speech alignment, acoustic modelling, and waveform synthesis in a single architecture. It integrates a conditional variational autoencoder (VAE) with normalizing flows and adversarial training, allowing it to directly generate waveforms without an intermediate vocoder stage. Text inputs are processed through a phoneme encoder, and monotonic alignment search is employed to align them with latent speech representations, eliminating the need for precomputed alignments. The decoder uses a HiFi-GAN–style adversarial waveform generator conditioned on latent variables sampled from the flow-based prior, enabling both high fidelity and controllable variation in speech output. By merging the traditionally separate stages of spectrogram generation and vocoding, VITS reduces pipeline complexity, improves prosody modelling, and achieves state-of-the-art naturalness in both objective and subjective evaluations.



\textbf{Grad-TTS}\footnote{\url{https://github.com/huawei-noah/Speech-Backbones/tree/main/Grad-TTS}}~\cite{popov2021gradttsdiffusionprobabilisticmodel} is a diffusion probabilistic model for text-to-speech synthesis that formulates Mel-spectrogram generation as a denoising process. Conditioned on text embeddings and predicted phoneme durations, the model iteratively refines Gaussian noise into a target spectrogram using a score-based model parametrised by a non-causal convolutional U-Net. The architecture employs monotonic alignment search for unsupervised text–speech alignment, avoiding the need for external aligners or attention mechanisms. A key advantage of Grad-TTS is the ability to control the trade-off between synthesis quality and inference speed by adjusting the number of denoising steps. This iterative refinement process allows for modelling fine-grained prosodic details and produces speech with high naturalness and expressiveness, rivalling both autoregressive and GAN-based systems, while retaining flexibility in controlling output characteristics.




\textbf{Matcha-TTS}\footnote{\url{https://github.com/shivammehta25/Matcha-TTS}}~\cite{mehta2024matcha} is a conditional flow-matching text-to-speech architecture that combines alignment learning, prosody modelling, and efficient generative modelling in a unified, end-to-end framework. The system learns to map text-conditioned latent representations to target Mel-spectrograms using conditional flow matching, an ODE-based generative process related to diffusion models but optimized for fast sampling. The architecture integrates a monotonic alignment search mechanism, enabling it to learn text–speech correspondences without external aligners, and uses a stochastic duration predictor to model prosodic variability. Compared to traditional diffusion-based TTS, Matcha-TTS achieves competitive perceptual quality with significantly fewer sampling steps, making it more suitable for low-latency scenarios. Its design offers a balance between the probabilistic expressiveness of iterative generative models and the efficiency demands of real-time deployment.

Table~\ref{tab:tts_comparison} shows an overview of the selected text-to-speech synthesis tools and their contributions. 

For all TTS systems, except VITS, we relied on the \textbf{HiFi-GAN} vocoder\footnote{The universal V1 checkpoint was used: \url{https://github.com/jik876/hifi-gan?tab=readme-ov-file#pretrained-model}}~\cite{hifigan} to transform the Mel-spectrograms into waveforms.

\begin{table*}[ht]
\centering
 \footnotesize

\caption{Comparison of selected open-source TTS architectures.}
\begin{tabularx}{\textwidth}{p{0.1\textwidth}p{0.1\textwidth}p{0.1\textwidth}
                            p{0.1\textwidth}p{0.1\textwidth}p{0.35\textwidth}}
\toprule
\textbf{Model} & \textbf{Architecture} & \textbf{Alignment} & \textbf{Prosody} & \textbf{Output} & \textbf{Main contributions} \\ \toprule

\textbf{FastPitch} & Feed-forward Transformer (non-autoregressive) & Duration predictor & Explicit F0 prediction & Mel-spectrogram & Parallel spectrogram prediction with explicit pitch conditioning for controllable prosody and fast inference. \\ \midrule

\textbf{VITS} & VAE + normalizing flows + GAN vocoder & Monotonic alignment search & Latent variable sampling & Waveform & End-to-end architecture unifying alignment, acoustic modeling, and vocoding; high-fidelity direct waveform generation with controllable variation. \\ \midrule

\textbf{Grad-TTS} & Diffusion probabilistic model (score-based) & Monotonic alignment search & Duration-based prosody control & Mel-spectrogram & Iterative denoising process enabling quality–speed trade-off; fine-grained prosody modeling without external aligners. \\ \midrule

\textbf{Matcha-TTS} & Conditional flow matching (ODE-based) & Monotonic alignment search & Stochastic duration modeling & Mel-spectrogram & Efficient iterative generation with flow matching, combining probabilistic expressiveness with fewer sampling steps for low-latency TTS. \\ \bottomrule

\end{tabularx}
\label{tab:tts_comparison}
\end{table*}

\subsection{Text processing}
\label{sec:text_processing}

The selected speech generation systems rely on distinct text processing procedures and symbol sets. 
VITS and Matcha-TTS include phoneme conversion using the Phonemizer package\footnote{\url{https://github.com/bootphon/phonemizer}} with the eSpeak-NG backend\footnote{\url{https://github.com/espeak-ng/espeak-ng}} in their text cleaners. 
FastPitch and Grad-TTS use code-based cleaners which simply lowercase text, expand abbreviations and convert characters to the ASCII counterparts, but do not include any phonetic processing. Their default symbol sets are limited to standard English alphabet. 

To ensure consistency across our trained models, we standardized the text processing pipeline by applying phonetic transcription with Phonemizer and eSpeak-NG and adapting the symbol sets to Romanian. For 
all TTS systems, this step was integrated directly into the text pipelines. As the transcripts from training and evaluation datasets were manually checked and normalised, no additional text pre-processing was carried out with the exception of lower-casing.

\subsection{Speech Data}

For TTS systems' training we selected the Romanian SWARA Speech Corpus \cite{stan_sped2017}.\footnote{\url{https://zenodo.org/records/15736789}} SWARA comprises 21 hours of high-quality Romanian read speech recordings, contributed by 17 non-professional speakers. Each speaker read aloud between 921 and 1493 prompts randomly selected from newspaper articles. The recordings were conducted in a semi-professional studio environment, utilizing a sound-proof booth and high-quality microphones. 
The audio data is sampled at 44.1 kHz with 16-bit resolution, and each utterance was manually segmented and checked for transcription errors.
A significant subset of SWARA, i.e. 16 hours, comprises 880 unique utterances read aloud by \textit{all the 17 speakers}.
This subset represents a valuable parallel spoken dataset, enabling cross-speaker analysis and controlled data selection, and we therefore use it as the core training data.

Compared to the original release of SWARA, an updated version of it added two new speakers (BEA and MAR) and removed a very noisy one, TIM. So our baseline SWARA dataset comprises 18 speakers. The common set of text prompts was retained from all speakers, making the training data parallel. This is not a necessary step, but it can alleviate imbalanced datasets with respect to the speaker representation. 
We use 16 speakers to train our baseline models. We refer to these models as \textit{eigen-voice models}. This terminology was very common in the HMM-based TTS era, and it refers to models which are trained on multiple speakers, but in which the speaker identity is not explicitly provided at the system's input. This, in theory, enables the system to extract relevant acoustic information without the need to specialise in the discrimination of the speaker identities. Nowadays, most of the speech generation architectures are trained on a wide range of speaker and acoustic conditions data, making the eigen-voices somewhat implicit. 

The other two speakers--BAS~(female) and SGS~(male)--were reserved for speaker adaptation and finetuning. To ensure a more realistic test of the systems' robustness, these speakers were selected based on automatically estimated naturalness scores using UTMOS~\cite{saeki2022utmosutokyosarulabvoicemoschallenge}, and obtained from a batch of randomly selected utterances. Both of them exhibited rather low recordings' quality, which would make them representative of real-world data. For each speaker, we construct two subsets of samples: one with 10 samples and one with 1000 samples. The 10-sample subset is designed to simulate extremely low-resource finetuning scenarios (47.82 seconds of speech for BAS and 39.73 seconds for SGS). The 1000-sample subset (approximately 60 minutes for BAS and 50 minutes for SGS), while much larger, still reflects a realistic amount of data that could be acquired or selected in practice. Given the parallel nature of the SWARA corpus, the two speaker subsets contain the same set of prompts--this minimizes the impact of the linguistic coverage in the adaptation process's direct comparison.

\subsection{Training Environment and Setup}

Each of the four TTS tools was trained from scratch using the default hyperparameters listed in their code repositories. 
To ensure consistency across models, all audio was resampled to $22,050$ Hz and the same phoneme-based text processing pipeline was applied.
Each baseline eigen-voice model was trained for $1.125mln$ iterations with a batch size of 16. No speaker identity was used to condition the output of these baseline models. Grad-TTS and Matcha-TTS were trained on a single Tesla V100-SXM2 GPU with 32GB VRAM. VITS and FastPitch were trained on a single Tesla T4 GPU with 16GB of VRAM.

For finetuning we used the two subsets of 10 and 1000 samples from each of the two held-out speakers, BAS and SGS. We sampled the finetuning process at three iteration counts: 120, 400 and 1000. For the 10 sample finetuning we used a batch size of 4, and a batch size of 16 was used for the 1000 sample subset.


\section{Evaluation}

The evaluation methodology assesses the selected TTS tools from two major perspectives: usability and output quality. 
For usability we focus on setup complexity, documentation quality and community support. While the quality of the synthesised speech is evaluated objectively and subjectively across several dimensions.

We should note that we did not run the English baseline training recipes because our focus was not on benchmarking performance in high-resource languages, but rather on adapting the framework to low-/mid-resource settings. English baselines are already extensively documented in the original recipe, and replicating them would not provide additional insights beyond what is already established in the literature.

\subsection{Usability assessment}

When developing a novel TTS system, the final selection of the adopted tool is also based on its ease of use. As a result, we first rate the TTS tools according to the following criteria:
\begin{itemize}
    \item \textbf{installation and dependencies:} the ease of setting up each model, including  
    the library dependency complexity and the availability of setup scripts (such as Docker or requirements files);
    \item \textbf{documentation quality:} the clarity and completeness of available instructions, including troubleshooting guides and example usage;
    \item \textbf{reproducibility:} the ease of replicating and adapting the setup, training and inference processes, including the necessary language-dependent modifications;
    \item \textbf{training and finetuning experience:} the ease of training/finetuning the model with custom datasets and the amount of effort mandated by data preprocessing;
    \item \textbf{community support:} the level of support available from the maintainers and the community, based on the number of open/closed GitHub issues, responsiveness and issue resolution rates.
\end{itemize}

\begin{table*}[t!]
    \centering
    \caption{Setup and usability assessment of the selected TTS systems.}
    \begin{tabularx}{\textwidth}{lXXXX}
        \toprule
        \textbf{Criteria} & \textbf{VITS} & \textbf{FastPitch} & \textbf{Grad-TTS} & \textbf{Matcha-TTS} \\
        \midrule
        Setup difficulty & Simple & Moderate & Simple & Simple \\
        Step-by-step examples & \checkmark & \checkmark & \checkmark & \checkmark \\
        Pretrained checkpoints & \checkmark & \checkmark & \checkmark & \checkmark \\
        Manual fixes needed & Minor & Moderate & Moderate & Minor \\
       Adaptations for Romanian & Text processor & Text processor, Training pipeline & Text processor & Text processor \\
        Community support & Moderate & Active & Moderate & Active \\
        \bottomrule
    \end{tabularx}
    \label{tab:tool}
\end{table*}

Table~\ref{tab:tool} summarises our findings and a detailed description is provided next:

\textbf{FastPitch} has a detailed setup process with structured documentation, especially when using the Docker container. The current version requires additional tools like Apex for mixed precision training, which can cause conflicts with some modules (CUDA, PyTorch). Additionally, specific GPU requirements (Volta, Turing and Ampere architectures) are needed for optimal performance. As a result we resorted to an older version (commit ID: \texttt{72a15ee}), which eliminates the Apex dependency. To ensure reproducibility on modern infrastructure, we established a stable environment using Python 3.12.3, PyTorch 2.6.0 and CUDA 12.4, running on NVIDIA Turing Architecture GPUs.
This older version also includes lighter and simpler training scripts. Adaptations for Romanian included modifying the text cleaner and phoneme processing pipeline, as mentioned in section \ref{sec:text_processing}. Setup was challenging initially, but after reverting to the aforementioned version, the training and finetuning proved intuitive. Community support seems to be active, with 289 open issues and 575 closed issues on GitHub, although recent maintenance involvement has been minimal.

\textbf{VITS} is easy to set up, with very brief instructions for installation and dependency management. However, it lacks a clear documentation on dataset format and training configuration files. This makes the adaptation process difficult for less experienced users. VITS allows for straightforward finetuning, but community support is limited, with 156 open and 57 closed issues, and maintainer engagement being minimal. It works out-of-the-box for English, but adapting it for languages like Romanian requires adjustments particularly in text preprocessing, which in this case is a separate preliminary step that cleans the original text before training.


The setup of \textbf{Grad-TTS} requires addressing specific implementation gaps to achieve full reproducibility. While the repository exhibits moderate official maintenance, evidenced by only 23 open and 11 closed issues, necessary solutions are readily available through community contributions. First, the compilation of the \texttt{monotonic\_align} module initially failed due to a missing directory in the source tree. Second, we integrated unmerged patches to stabilize training: the text encoder was updated to ensure proper usage of embedding information for the multi-speaker configuration, and the mel-spectrogram cropping logic was corrected to prevent shape mismatch failures when processing short utterances. Finally, adapting the model to Romanian required replacing the original CMUDICT-based phoneme extraction with eSpeak to generate compatible phoneme IDs.

\textbf{Matcha-TTS} offers a straightforward setup process accompanied by comprehensive documentation and numerous command line tools for interaction. Its design facilitates easy adaptation to new datasets through clear, detailed instructions covering configuration and feature extraction pipelines. It includes readily available example configurations for both single and multi-speaker systems, alongside flexible training options such as multi-GPU support and neatly organized experiments' configurations. 
Adapting it to Romanian primarily involved executing the feature extraction script, adjusting the text processing pipeline for the language, and modifying the experiment configuration to utilize our specific dataset and extracted features. The project's GitHub repository indicates active community support with 18 open and 93 closed issues. User queries are timely addressed, although active development appears to have slowed down lately.

Among the evaluated systems, \textbf{Matcha-TTS} stands out as the most user-friendly, providing the most comprehensive documentation, a structured setup process and flexible adaptation pipeline. \textbf{VITS} is lightweight and simple to install, yet its limited documentation and separate text processing step may represent an obstacle for less experienced users. \textbf{FastPitch} can seem less accessible due to its heavier setup requirements and more involved text processing adaptations. \textbf{Grad-TTS} is one of the least favoured options, as the repository required fixes and introduced additional overhead when running experiments for languages other than English.

\subsection{Objective Assessment of TTS Output Quality}

Objective measures of TTS quality have recently gained more academic and commercial interest as complementary assessments of a system's output. They can, to some extend, estimate the intelligibility, naturalness and speaker similarity without relying on subjective listening tests. 

On top of tool-level evaluation, we employ a finer-grained analysis of speech quality influencing factors, like adaptation data size and compute time, or speaker characteristics (here, male vs female). Each metric is computed per utterance, and the average over the 42 synthesized evaluation samples (held-out set) is reported. Results for the the speaker adaptation using 10 or 1000 speech samples, and varying the number of adaptation iterations (120, 400 or 4000) for the male and female speakers are also discussed.

In order to perform this objective assessment, we resort to the extended set of measures described in the following subsections. Although not always in line with the results obtained using listening tests, this extended set of metrics allows for a comprehensive and reproducible assessment of TTS performance.

\subsubsection{Pronunciation accuracy and intelligibility assessment}
Word Error Rate (\textbf{WER}) and Character Error Rate (\textbf{CER}) are used to assess pronunciation accuracy and intelligibility. Both metrics were computed by comparing Whisper-transcribed \cite{radford2022robustspeechrecognitionlargescale} outputs 
with the corresponding reference transcripts using the Jiwer\footnote{\url{https://jitsi.github.io/jiwer/usage/}} library. 
The same text normalization process was applied to both transcriptions, including lowercasing, removing punctuation, mapping common abbreviations to their full forms, and converting numbers into words.

\begin{figure*}[t!]
    \centering
    \includegraphics[trim=0 0 0 0, clip, width=0.8\linewidth]{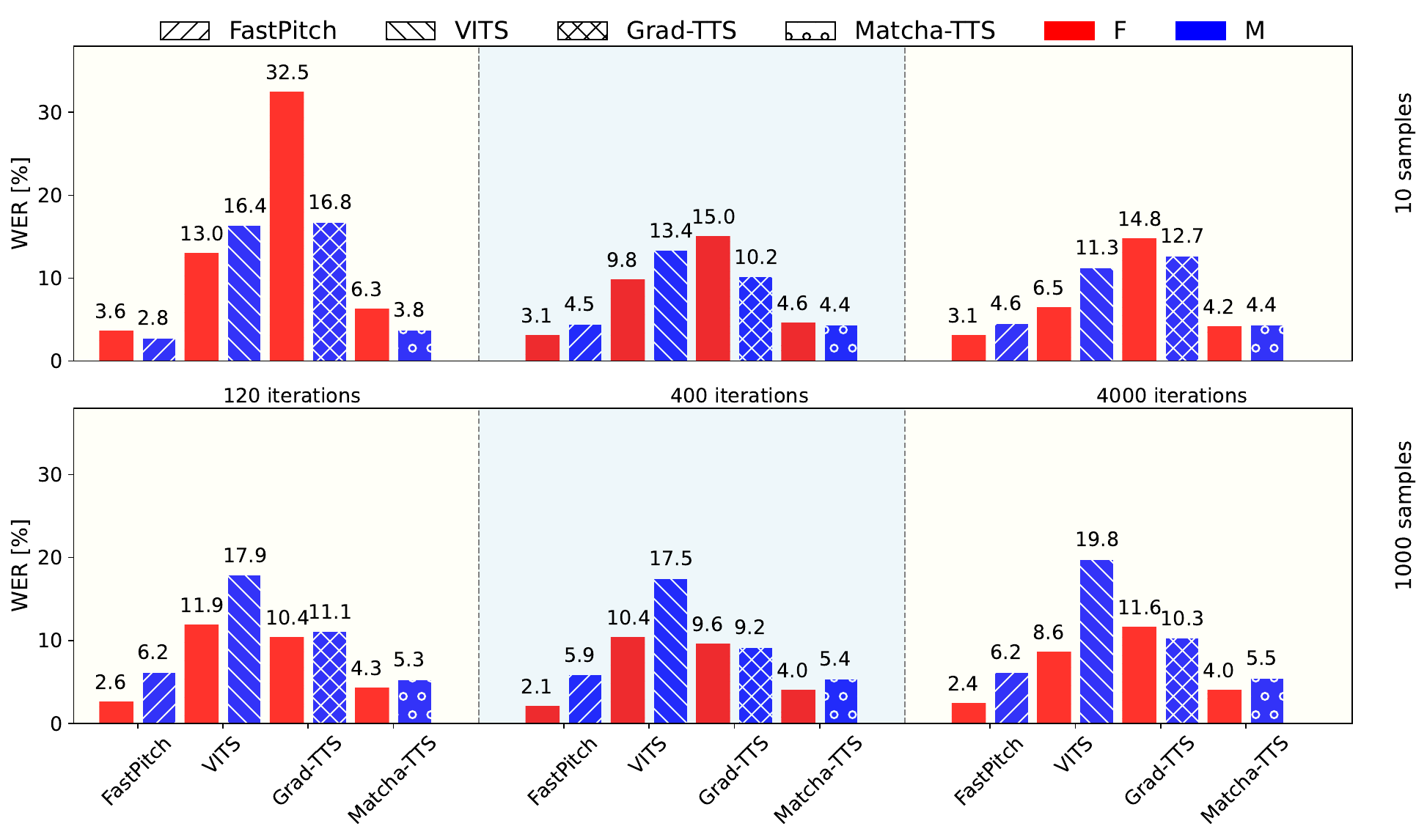}
    \caption{%
       Pronunciation accuracy and intelligibility, reported as Word Error Rate - WER ($\downarrow$) scores, for all models finetuned with 10 utterances per speaker (top) and 1000 utterances per speaker (bottom), evaluated at different stages of the finetuning process: 120 iterations (left); 400 iterations (center), and 4000 iterations (right).}
    \label{fig:wer}
\end{figure*}

Figure~\ref{fig:wer} reveals clear differences among the intelligibility of the models. FastPitch achieved the lowest error rates across all finetuning conditions, with the best result of 2.1\% WER for the female speaker using 1000 training samples. Matcha-TTS followed closely and maintained a stable performance across both low and high resource scenarios--its WER consistently remained below 6\% for both speakers, suggesting robustness to data volumes.
In contrast, VITS and Grad-TTS showed weaker performances. Grad-TTS produced the highest WERs, although its accuracy improved substantially with more training (from 32\% to 14\% for the female speaker). VITS displayed instability for the male speaker, where additional training led to a slight degradation (from 17\% to 19\% WER). Manual inspection confirmed that VITS often generated truncated utterances, which contributed to the high WER.
Across all systems, adaptation to the female voice resulted in lower WER. Yet, this is correlated to the natural speech samples, where the female speaker yielded a 3\% WER compared to 7\% for the male.

\subsubsection{Naturalness assessment}
Mean Opinion Score (\textbf{MOS}) based on the UTMOS automatic prediction system~\cite{saeki2022utmosutokyosarulabvoicemoschallenge} is used to assess perceived naturalness. This was evaluated with the implementation provided in the Discrete Speech Metrics package.\footnote{\url{https://github.com/Takaaki-Saeki/DiscreteSpeechMetrics}}
Test waveforms were resampled to 16 kHz before applying the predictor to obtain a non-intrusive MOS estimate. This automatic approach was selected as a scalable and reproducible alternative to subjective listening tests.

\begin{figure*}[t!]
    \centering
    \includegraphics[trim=0 0 0 0, clip, width=0.8\linewidth]{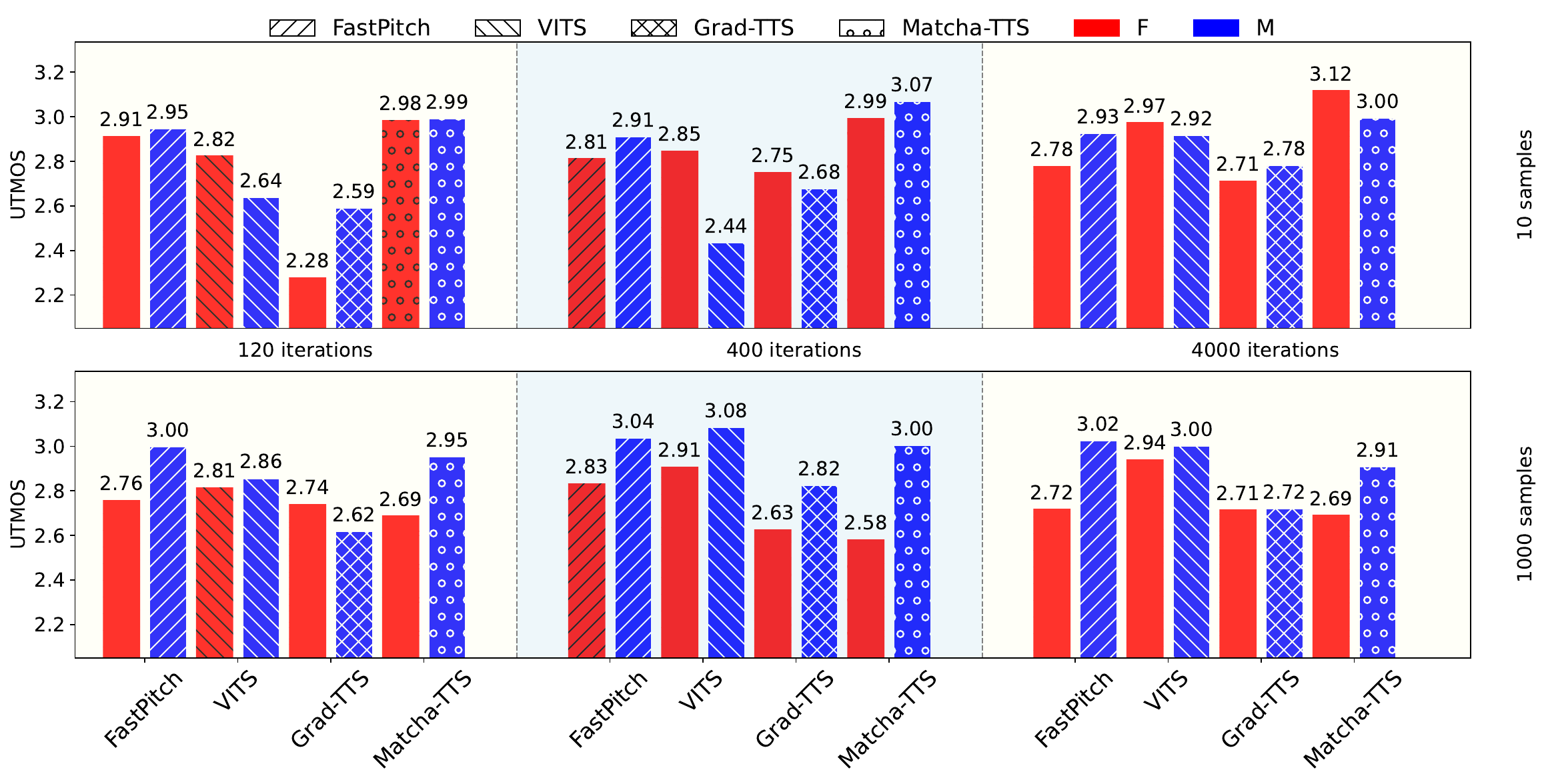}
    \caption{%
       Perceived naturalness, reported as Mean Opinion Score - UTMOS ($\uparrow$) scores, for all models finetuned with 10 utterances per speaker (top) and 1000 utterances per speaker (bottom), evaluated at different stages of the finetuning process: 120 iterations (left); 400 iterations (center), and 4000 iterations (right).
       }
    
    \label{fig:utmos}
\end{figure*}

\textbf{UTMOS} scores (shown in Figure~\ref{fig:utmos} 
are automatic evaluations of the perceived naturalness of speech, with higher values indicating more natural-sounding synthesis. Baseline scores from the original recordings were $3.07$ for the female speaker and $2.92$ for the male speaker. These values provide a reference for the assessment of the model performance.
Across all systems, UTMOS scores ranged from $2.28$ to $3.12$, with the male speaker generally perceived as slightly more natural, despite the female speaker achieving lower WER and better speaker similarity. Matcha-TTS and FastPitch consistently delivered the best results, while VITS performed competitively in the 1000-sample scenarios, occasionally surpassing both. However, VITS lacked stability: in the 10-sample case, its scores fluctuated thus reflecting instability in its generation process. 
FastPitch remained steady across training conditions, peaking at $3.02$ in the 1000-sample setup. Matcha-TTS consistently scored high but showed an unexpected dip for the female speaker in the 1000-sample scenario. Grad-TTS, while improving with more training data, generally remained below the other models, achieving competitive results only in the high-resource scenario. It should be noted that this objective assessment of naturalness is not always correlated with the subjective assessment, performed with listening tests (see section \ref{sec:subjective_evaluation}).

\subsubsection{Speaker similarity assessment}
Speaker Encoder Cosine Similarity (\textbf{SECS}) using the Voice Encoder~\cite{8462665} from the Resemblyzer\footnote{\url{https://github.com/resemble-ai/Resemblyzer}} package quantifies speaker identity preservation by computing the cosine similarity between embeddings extracted from each synthesized utterance and its paired reference recording. Prior to embedding, audio was converted to mono and resampled to 16 kHz.

\begin{figure*}[t!]
    \centering
    \includegraphics[trim=0 0 0 0, clip, width=0.8\linewidth]{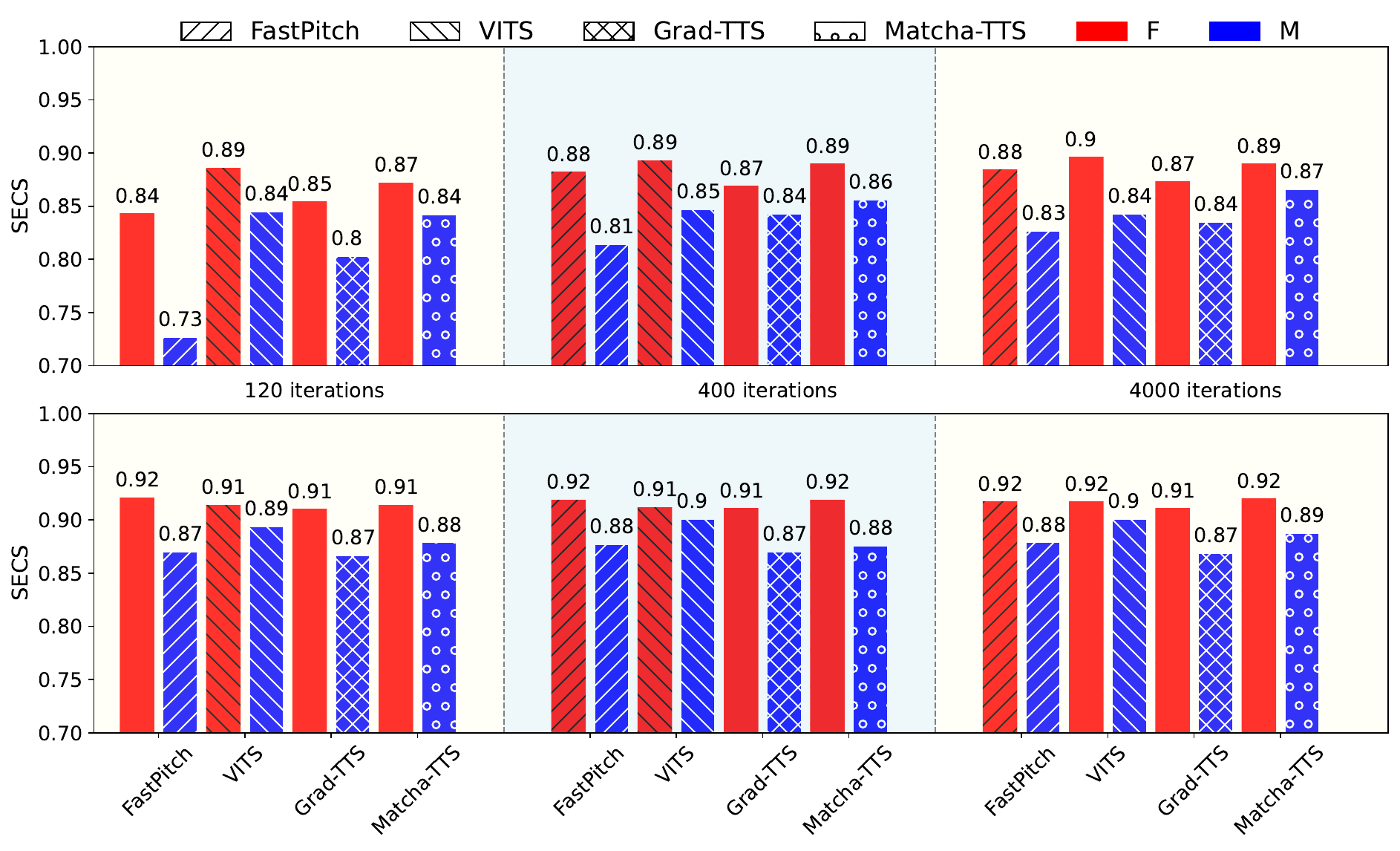}
    \caption{%
       Speaker similarity, reported as Speaker Encoder Cosine Similarity - SECS ($\uparrow$) scores, for all models finetuned with 10 utterances per speaker (top) and 1000 utterances per speaker (bottom), evaluated at different stages of the finetuning process: 120 iterations (left); 400 iterations (center), and 4000 iterations (right). 
       }
    \label{fig:secs}
\end{figure*}

High \textbf{SECS} scores are desirable, indicating better replication
of speaker identity. Figure~\ref{fig:secs} shows the results for the four architectures, and we can draw the following insights: 
speaker cosine similarity increases consistently with more training data: models trained on 1000 samples outperform those trained on only 10, while additional finetuning iterations have a lower impact. Across all systems, the female speaker yields higher similarity scores than the male one.
VITS achieves the highest speaker similarity overall (peaking at $0.92$), preserving identity even in low-data settings, while FastPitch improves with more samples, particularly for the male speaker. Matcha-TTS delivers competitive and stable results across all training scenarios, though slightly below VITS. Grad-TTS shows the lowest similarity scores, improving with data but remaining weaker than the other models.
Taken together, these results suggest a trade-off between intelligibility and identity preservation: FastPitch leads in WER, but VITS more faithfully captures speaker characteristics. Matcha-TTS seemingly provides a more balanced performance.

\subsubsection{Perceptual quality, intelligibility, and signal fidelity assessment}
We also adopt Perceptual Evaluation of Speech Quality (\textbf{PESQ}), Short-Time Objective Intelligibility (\textbf{STOI}), and Scale-invariant signal-to-distortion ratio (\textbf{SI-SDR}) metrics stemming from the TorchAudio-Squim estimator~\cite{10096680}.
These non-intrusive metrics provide complementary measures of perceptual quality, intelligibility, and signal fidelity.
In this case, the audio was resampled to the model’s expected sampling rate.\footnote{\url{https://docs.pytorch.org/audio/2.6.0/generated/torchaudio.pipelines.SQUIM_OBJECTIVE.html}}

\begin{figure*}[t!]
    \centering
    \includegraphics[trim=0 0 0 0, clip, width=0.8\linewidth]{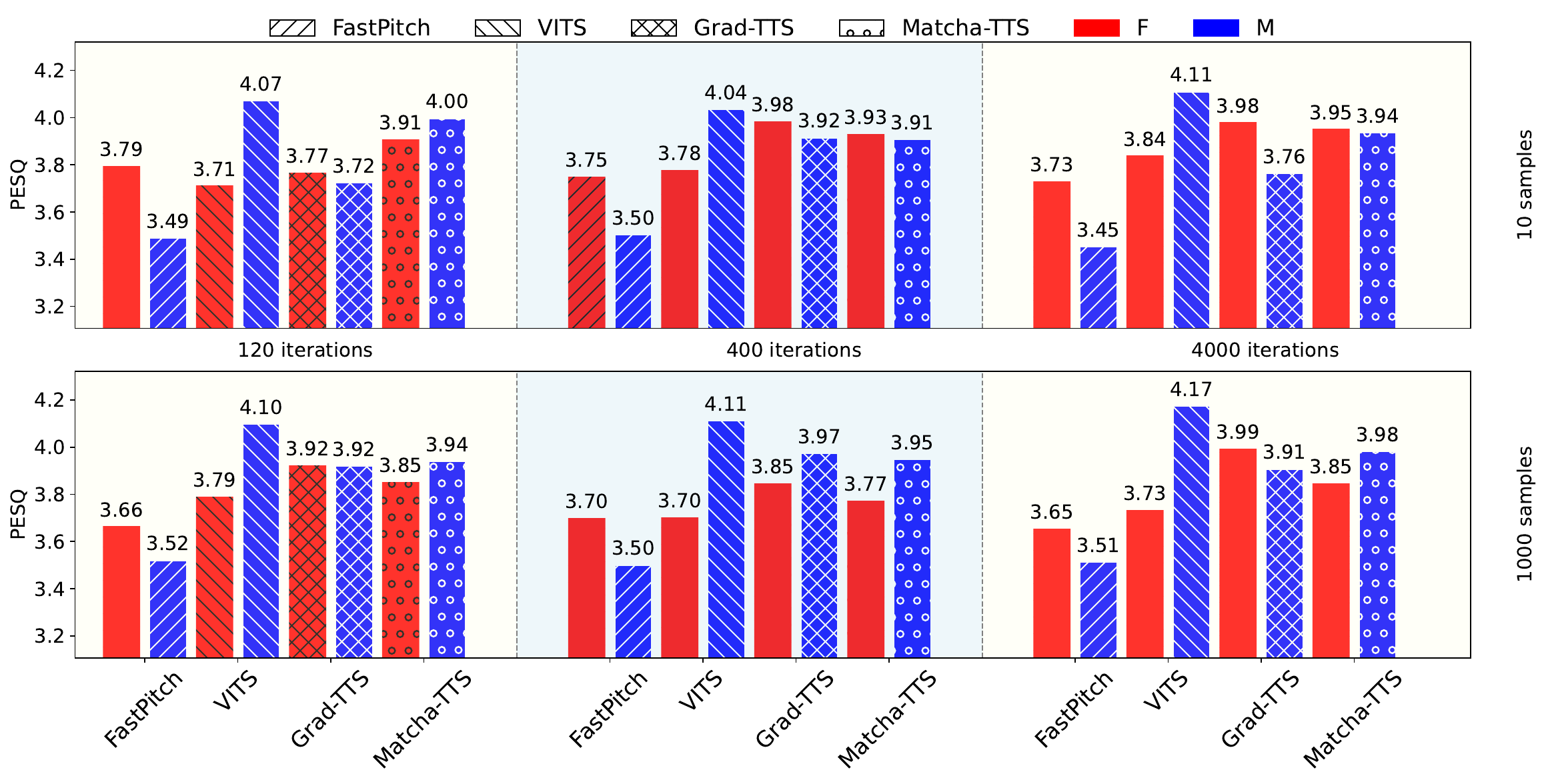}
    \caption{%
       Perceptual quality, reported as Perceptual Evaluation of Speech Quality - PESQ ($\uparrow$) scores, for all models finetuned with 10 utterances per speaker (top) and 1000 utterances per speaker (bottom), evaluated at different stages of the finetuning process: 120 iterations (left); 400 iterations (center), and 4000 iterations (right).
       }
    \label{fig:pesq}
\end{figure*}

\begin{figure*}[t!]
    \centering
    \includegraphics[trim=0 0 0 0, clip, width=0.8\linewidth]{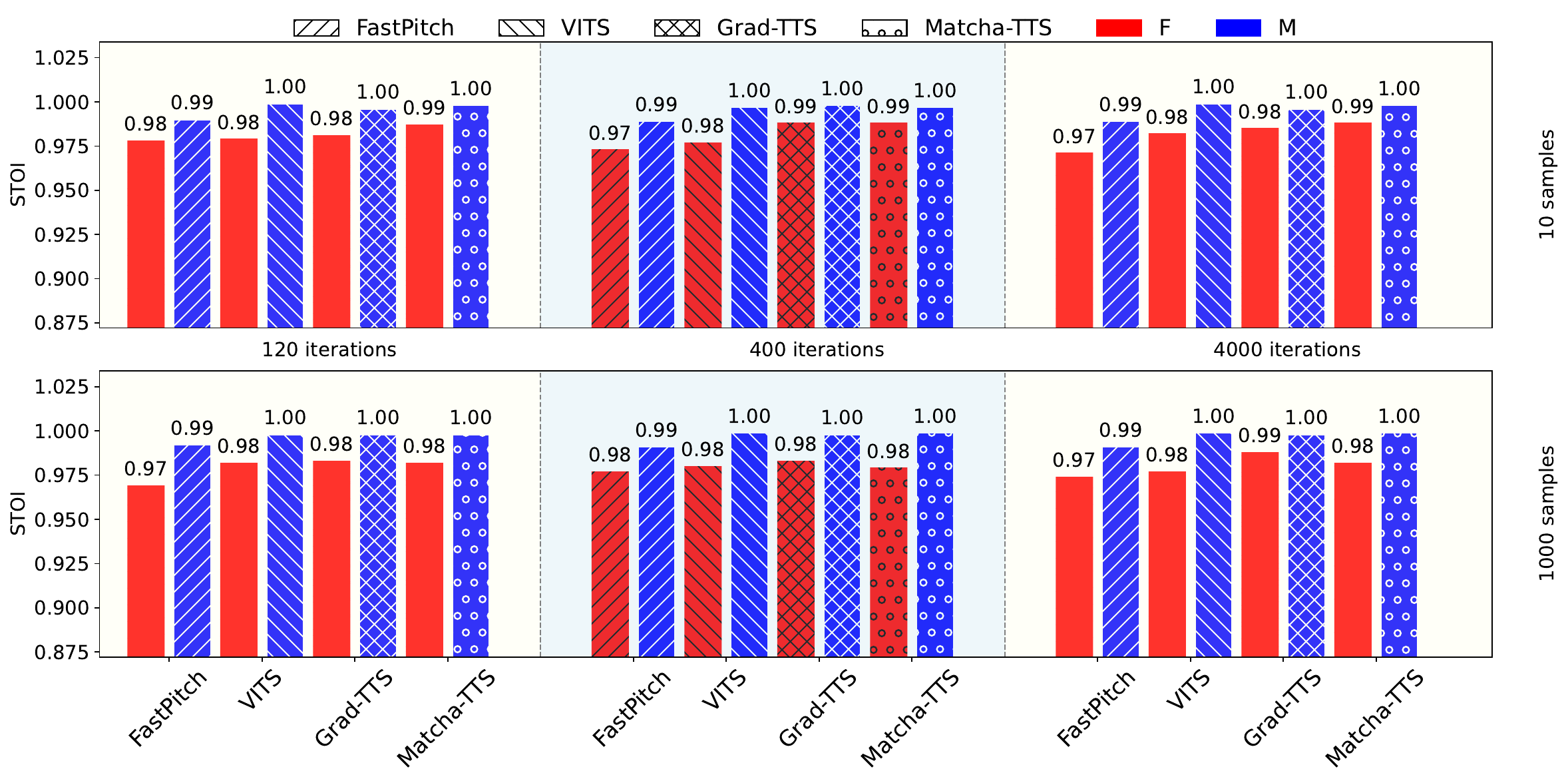}
    \caption{%
       Perceptual inteligibility, reported as Short-Time Objective Intelligibility - STOI ($\uparrow$) scores, for all models finetuned with 10 utterances per speaker (top) and 1000 utterances per speaker (bottom), evaluated at different stages of the finetuning process: 120 iterations (left); 400 iterations (center), and 4000 iterations (right).
        }
    \label{fig:stoi}
\end{figure*}

\begin{figure*}[t!]
    \centering
    \includegraphics[trim=0 0 0 0, clip, width=0.8\linewidth]{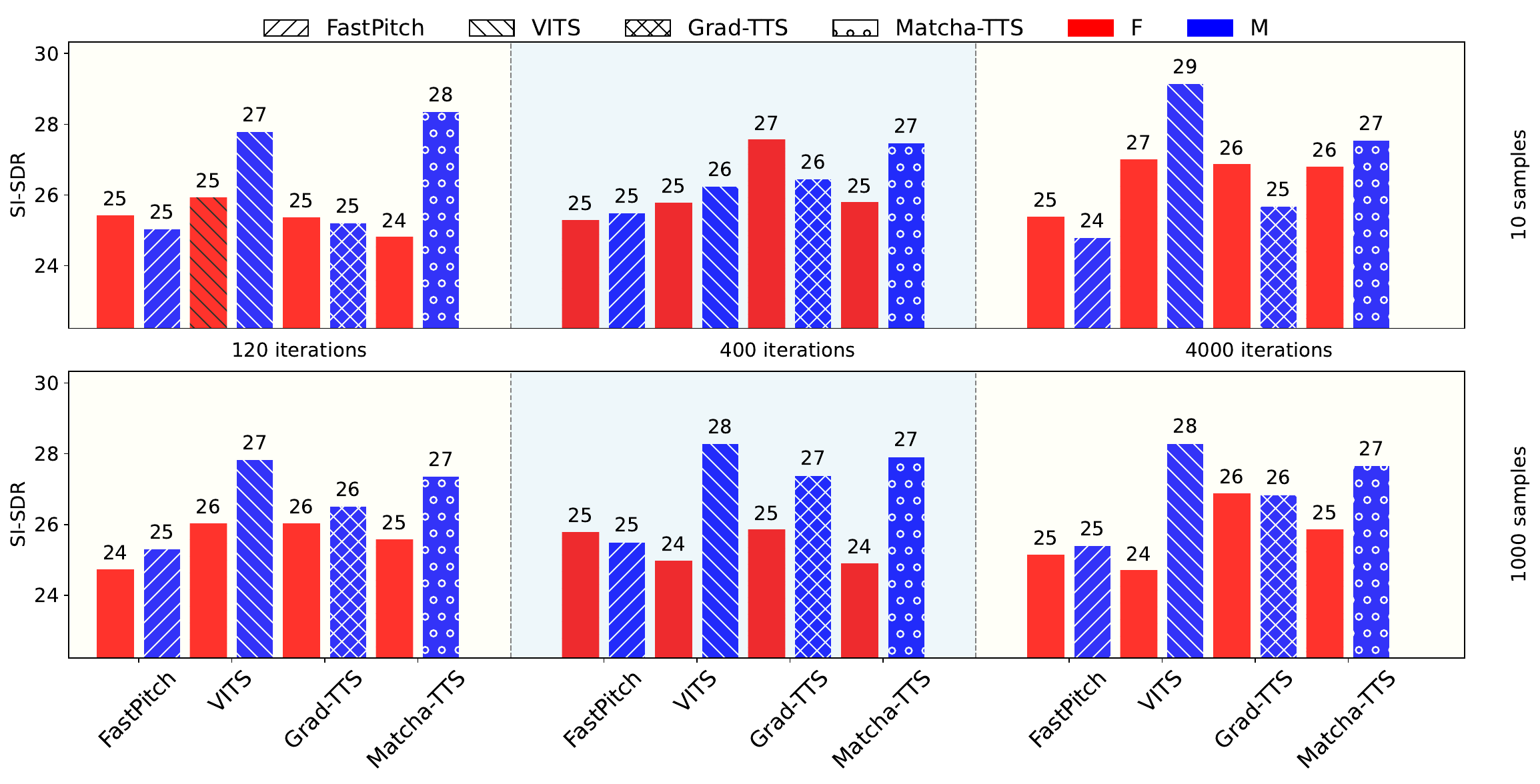}
    \caption{%
       Signal fidelity, reported as Scale-invariant signal-to-distortion ratio - SI-SDR ($\uparrow$) scores, for all models finetuned with 10 utterances per speaker (top) and 1000 utterances per speaker (bottom), evaluated at different stages of the finetuning process: 120 iterations (left); 400 iterations (center), and 4000 iterations (right).
    }
    \label{fig:sdr}
\end{figure*}

Across \textbf{PESQ} (Figure~\ref{fig:pesq}), \textbf{STOI} (Figure~\ref{fig:stoi}) and \textbf{SI-SDR} (Figure~\ref{fig:sdr}), the general trend is rather clear: all models produce intelligible speech of reasonable quality, but some handle perceptual fidelity better than others. Increasing training data from 10 to 1000 samples and extending iterations had only minor effects. 
VITS and Matcha-TTS consistently lead, meaning they generate speech that sounds more natural and less distorted. FastPitch and Grad-TTS are slightly weaker, indicating that perceptual quality and signal fidelity do not always align with improvements in WER or speaker similarity: a model can produce accurate pronunciations or succeed in preserving speaker identity while being perceptually less convincing.
Slight advantages for the male speaker show that speaker characteristics can influence perceptual and distortion-sensitive metrics, but differences are small.

\subsubsection{Prosodic and spectral deviation assessment}
Prosodic and spectral aspects were evaluated through the root mean squared error (\textbf{RMSE}) of the log F0 patterns and Mel Cepstral Distortion (\textbf{MCD}). 
Log F0 RMSE measures average differences in pitch contour between generated and reference speech extracted with a WORLD-based procedure over $40-800 Hz$ search range. Trajectories are converted to the logarithmic scale and aligned frame-wise before computing the per-utterance RMSE.
On the other hand, MCD quantifies spectral deviation of the synthesised waveform through Mel-frequency cepstral coefficients of order 25. 
Both metrics were evaluated with the DiscreteSpeechMetrics package using a window size of 512 and a hop of 256 samples.

\begin{figure*}[t!]
    \centering
    \includegraphics[trim=0 0 0 0, clip, width=0.8\linewidth]{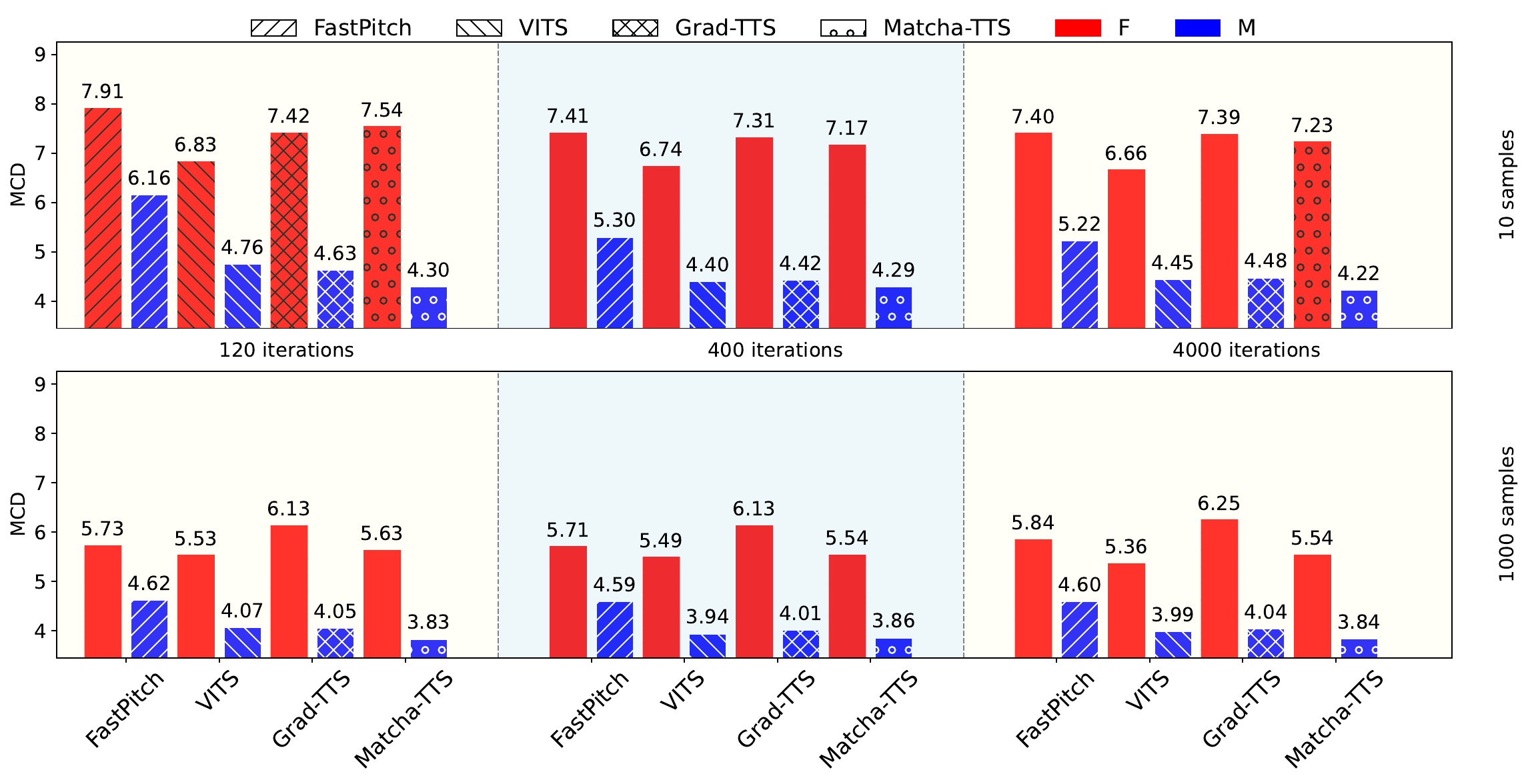}
    \caption{%
       Spectral deviation of the synthesised waveform, reported as Mel Cepstral Distortion - MCD ($\downarrow$) scores, for all models finetuned with 10 utterances per speaker (top) and 1000 utterances per speaker (bottom), evaluated at different stages of the finetuning process: 120 iterations (left); 400 iterations (center), and 4000 iterations (right).
    }
    \label{fig:mcd}
\end{figure*}

\textbf{MCD} (shown in Figure~\ref{fig:mcd}) quantifies spectral differences between natural and synthesized speech, with lower values indicating closer alignment. Across all models, MCD was consistently better for the male speaker, contrasting with other metrics like WER or Log F0 RMSE where the female speaker typically performed better. Increasing training data had the largest impact: the 10-sample scenario yielded a minimum of $4.22$, while 1000 samples reduced it to $3.83$. Additional training iterations contributed only marginal improvements. Model-wise, Matcha-TTS consistently achieved the best scores, peaking at $3.83$, while VITS excelled in the 10-sample scenario and remained competitive at higher resources. FastPitch struggled under little adaptation data conditions but improved with more data, and Grad-TTS remained the weakest overall. Overall, spectral fidelity benefits primarily from larger datasets, with Matcha-TTS and VITS showing the strongest ability to reproduce the natural spectrum.

\begin{figure*}[t!]
    \centering
    \includegraphics[trim=0 0 0 0, clip, width=0.8\linewidth]{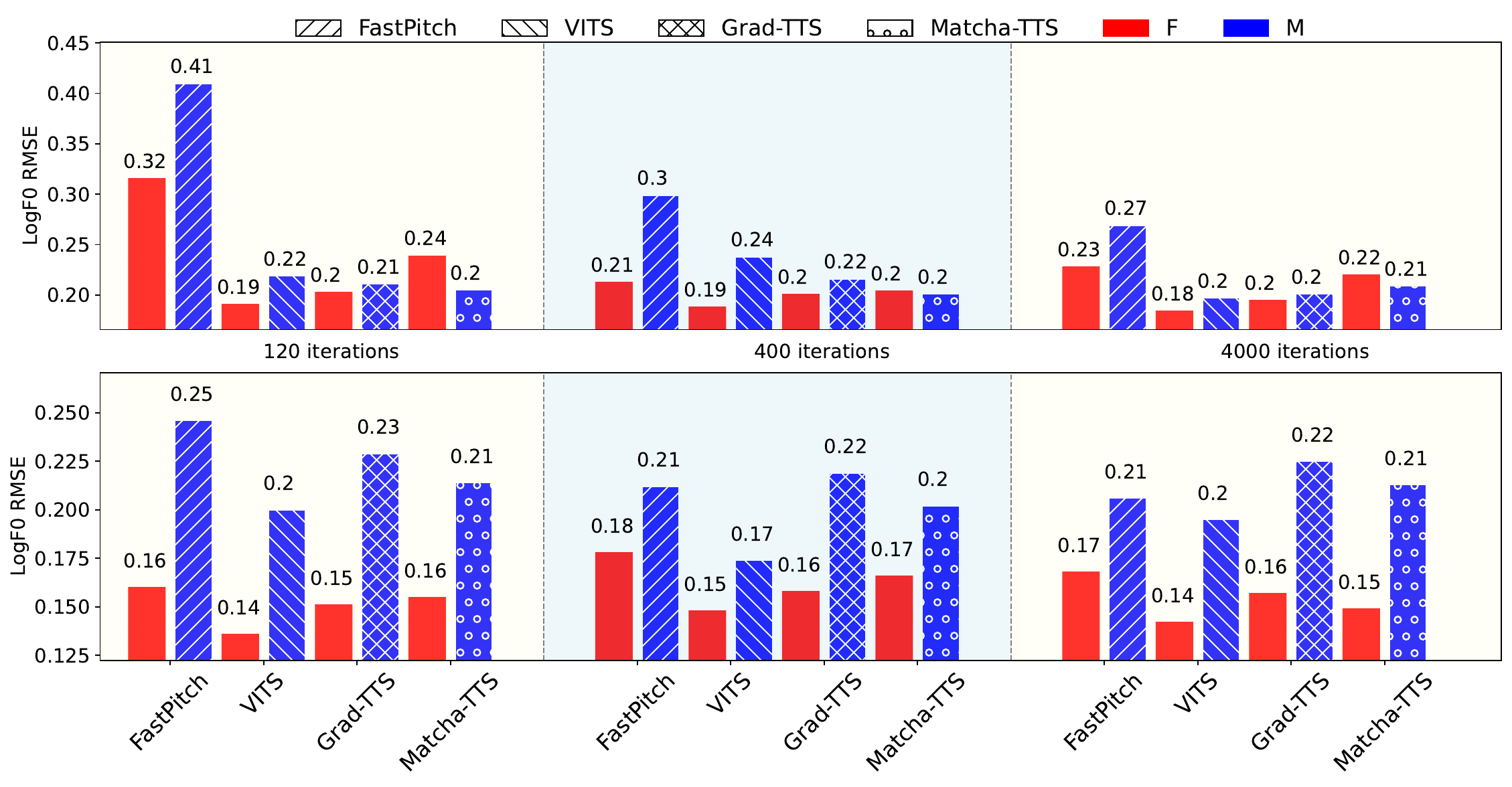}
    \caption{%
       Differences in pitch contour, reported as Log F0 RMSE ($\downarrow$) scores, for all models finetuned with 10 utterances per speaker (top) and 1000 utterances per speaker (bottom), evaluated at different stages of the finetuning process: 120 iterations (left); 400 iterations (center), and 4000 iterations (right).
    }
    \label{fig:logf0}
\end{figure*}

\textbf{Log F0 RMSE} quantifies deviations in pitch contour between synthesized and reference speech, with lower values indicating closer alignment and more accurate pitch reproduction. The results are summarized in Figure~\ref{fig:logf0} with scores ranging from $0.14$ to $0.41$.
All models achieved their best performance for the female speaker in the 1000-sample scenario, with minimum RMSE values at $0.14$, demonstrating that ample training data allows models to capture fine-grained pitch variations. For the male speaker, however, additional training data did not improve pitch accuracy, suggesting that male pitch contours were inherently more challenging to model under these settings.
Across iterations, log F0 RMSE was generally stable in both the 10 and 1000 samples scenarios, except for FastPitch in the 10-sample scenario, which improved substantially with more training iterations (from $0.41$ to $0.27$ for the male speaker).
FastPitch struggled under little adaptation data conditions but improved steadily with more data, Matcha-TTS was consistently stable, VITS achieved the overall best score, and Grad-TTS showed relative strength in pitch despite weaker WER and speaker similarity. Overall, pitch reproduction benefits from sufficient training data, female voices are easier to model, and FastPitch is the least stable under limited resources.

\subsubsection{Summary of objective assessment}
Overall, the objective results highlight that once a model has enough data to capture a speaker's style, further training contributes only marginally to improvements in perceptual quality. Instead, choosing the right architecture has a larger impact than simply adding training data or allowing the model to learn for more iterations.
FastPitch and Matcha-TTS perform well even in little data conditions. For instance, FastPitch achieves a WER of $0.03$, a SECS of $0.72$ and an UTMOS of $2.94$ for the male speaker using only 10 training samples and 120 finetuning steps, matching or surpassing baseline scores. Similarly, Matcha-TTS reaches a WER of $0.05$, a SECS score of $0.87$ and a UTMOS of $2.98$ for the female speaker under comparable conditions. In contrast, Grad-TTS and VITS require more data and training to achieve competitive performance.

\subsection{Subjective Evaluation of TTS Output Quality}
\label{sec:subjective_evaluation}

A clearer picture of the synthesis quality can be obtained through subjective listening tests. We opt for naturalness and speaker similarity evaluations for the two target speaker voices, BAS and SGS. A group of 31 listeners were asked to rate the TTS models finetuned with 1000 samples for 4000 iterations. Separate listening tests were set up for the two speakers. The samples were rated on a 0-unnatural to 100-natural scale, and all systems were presented to the listeners in parallel. The corresponding natural sample was mixed among the synthetic samples. For the speaker similarity test, the natural sample was clearly marked in the page. The test was carried out by the listeners in their home environments, but they were asked to listen over headphones. 
The instructions were: 
\begin{itemize}
    \item \textit{``Assign a score from 0-bad to 100-excelent to evaluate the overall quality of the voice. Focus on how natural and intelligible each sample sounds''} and
    \item \textit{``Assign a score from 0-bad to 100-excellent to evaluate the similarity of the synthesized voice to the original speaker. The original speaker sample can be accessed by pressing the 'Reference' button located on the left side of the test page.''} 
\end{itemize}

\begin{figure}[t!]
    \centering
    \includegraphics[trim=0 0 0 0, clip, width=\linewidth]{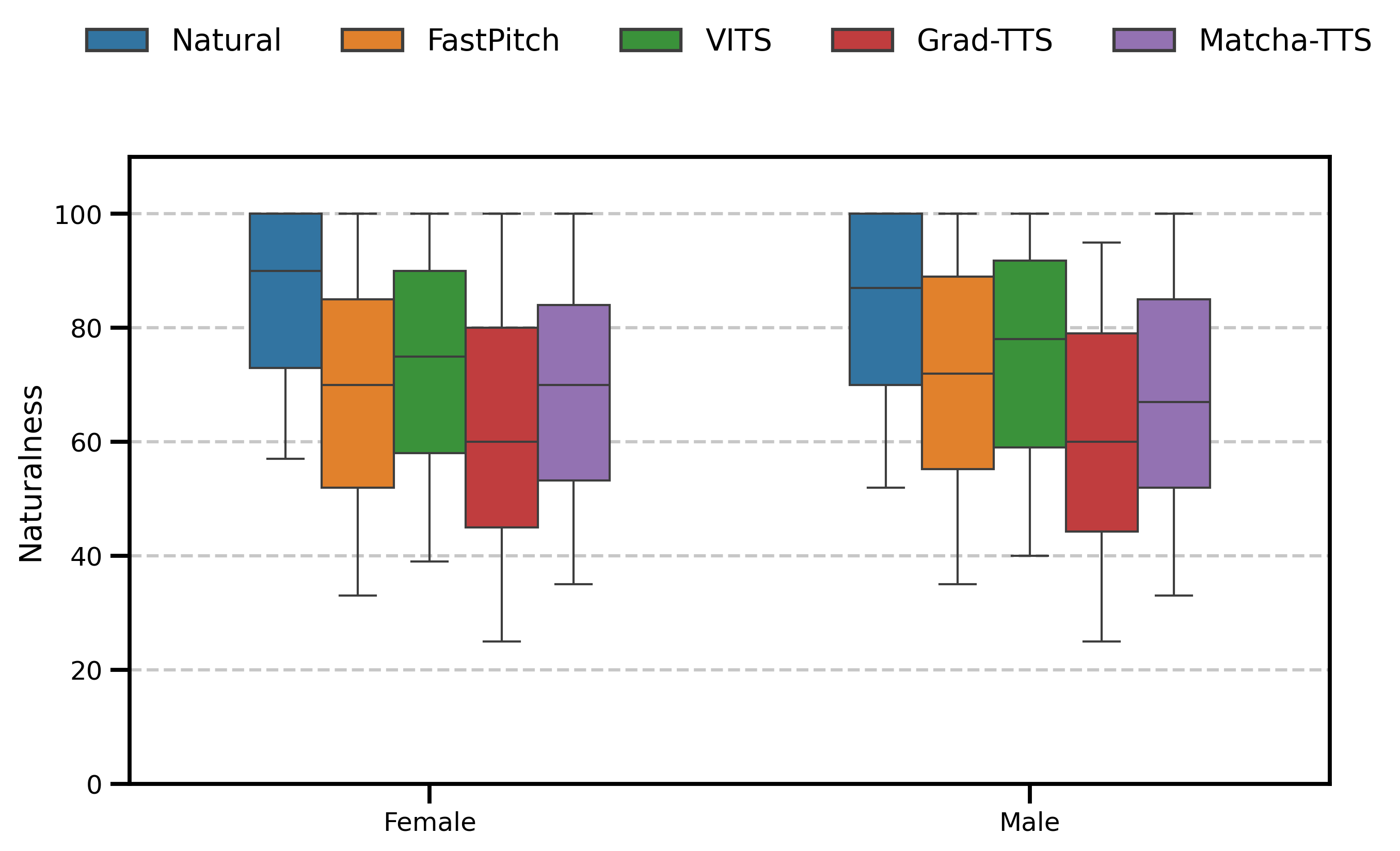}
    \caption{%
       Perceived Naturalness score statistics obtained in the listening test for the models finetuned with 1000 utterances per speaker for 4000 iterations.
    }
    \label{fig:mos}
\end{figure}

\begin{figure}[t!]
    \centering
    \includegraphics[trim=0 0 0 0, clip, width=1.\linewidth]{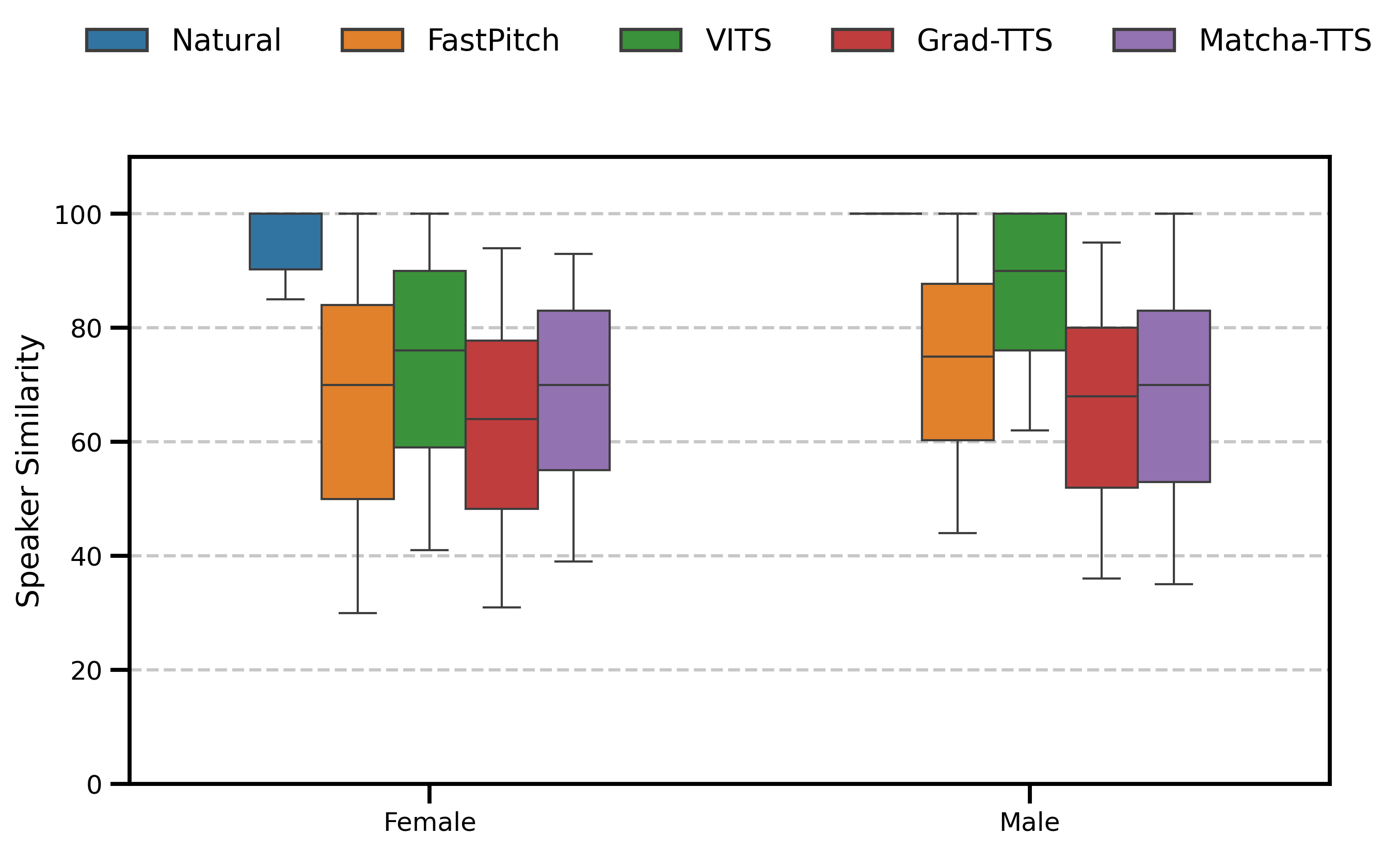}
    \caption{%
       Perceived Speaker Similarity score statistics obtained in the listening test for the models finetuned with 1000 utterances per speaker for 4000 iterations.
    }
    \label{fig:spk}
\end{figure}

Results of the listening test are shown in Figures~\ref{fig:mos}~and~\ref{fig:spk}. Two listeners were dropped due to high score variance and flat rating.  We can notice that VITS exhibits best performance in naturalness and speaker similarity, for both speakers, with FastPitch being consistently rated as second-best option. 

The listener ratings were also analysed using one-way ANOVA followed by Tukey HSD post-hoc tests to assess statistical differences among the speech synthesis systems. For naturalness ratings, FastPitch and Matcha-TTS achieved comparable scores, showing no significant difference from each other (p > 0.05). Natural, Grad-TTS, and VITS each differed significantly from the other systems, indicating substantial variability in perceived naturalness.
Speaker similarity ratings showed a similar pattern. FastPitch and Matcha-TTS were again comparable, with no significant difference between them. 

\section{Discussions and Conclusions}

This study explored the adaptability of four open-source TTS models to a new language. 
To bridge compatibility with English-oriented models, we adopted a language-agnostic text processing pipeline. Unlike other surveys on speech synthesis systems that focus primarily on reported results, we approached these models from a practical user perspective. We evaluated not only their output quality, but also the ease of setup and training in uncovered languages, including speaker adaptation experiments to test generalization.

Across our evaluations, FastPitch consistently demonstrated strong intelligibility, achieving the lowest WER in all experimental conditions. Regarding naturalness, FastPitch ranked second in the listening tests and placed first or second in most objective evaluation scenarios. For speaker similarity, both the objective and subjective assessments positioned FastPitch as the second-best system, following VITS.

VITS achieved the highest scores in all subjective evaluations, indicating clear listener preference. However, its objective intelligibility performance was substantially weaker, consistently ranking third or fourth among the tested systems, with word error rates frequently exceeding 10\%. Notably, none of the 31 listeners explicitly reported intelligibility issues, although our protocol did not directly solicit such feedback. We hypothesize that VITS-generated speech may contain artifacts that interfere with Whisper-based transcription, thereby inflating its WER. In terms of speaker similarity, VITS consistently ranked first, performing comparably to FastPitch and Matcha-TTS in the objective assessment and outperforming all systems in the subjective evaluations.

Matcha-TTS achieved the highest naturalness scores in the objective evaluation according to UTMOS. However, human listeners did not concur, ranking Matcha-TTS third in both naturalness and speaker similarity. This discrepancy illustrates a clear limitation of objective naturalness metrics such as UTMOS, whose predictions did not align with human perceptual judgments. Nevertheless, Matcha-TTS showed strong robustness in the objective evaluation, maintaining competitive performance even with minimal finetuning data. In contrast, VITS and Grad-TTS generally required larger amounts of data and additional training effort to reach similar performance levels.

Grad-TTS typically ranked last in naturalness across both objective and subjective evaluations. Its intelligibility performance was similarly limited, with WER values often exceeding 10\% and placing it third or fourth among the systems. As with VITS, it is plausible that Grad-TTS outputs contain artifacts that Whisper struggles to transcribe reliably.

From a user experience perspective, Matcha-TTS was the most accessible TTS system, offering clear instructions and tooling for customization. FastPitch, while powerful and well documented, required careful requirement setup due to module incompatibilities, which could be a barrier for some users. Grad-TTS and VITS offered less guidance, with Grad-TTS depending heavily on community fixes.

In preparation for this study, we also explored the adaptation of Parler-TTS~\cite{lyth2024parlertts}, which incorporates not only text prompts, but also structured descriptions of speech attributes and recording conditions. Parler-TTS proved difficult to adapt, revealing the fragility of state-of-the-art TTS pipelines when used with non-English data. Even after implementing a multi-stage workaround for generating Romanian speech attribute descriptors, the model failed to train from scratch and produced only silence. Although fine-tuning an existing checkpoint worked, it diverged from our intended protocol (i.e. train from scratch), so Parler-TTS was ultimately excluded.


We believe that this work serves as a practical reference for researchers and practitioners aiming to adapt TTS models to languages beyond their original design, and contributes to the development of more inclusive and accessible speech technologies.

\section*{Acknowledgment}

This work was co-funded by EU Horizon project AI4TRUST (No. 101070190) and by a grant of the Ministry of Research, Innovation and Digitization, CCCDI - UEFISCDI, project number PN-IV-P7-7.1-PTE-2024-0546, within PNCDI IV.


\bibliographystyle{IEEEtran}
\bibliography{access}

\end{document}